\documentclass[10pt,twoside,BCOR7mm,DIV15,headinclude,footexclude,cleardoubleempty,idxtotoc]{scrartcl}

\usepackage[english]{babel}
\usepackage{graphicx}
\usepackage{hyperref}
\usepackage{scrpage2}
\usepackage{hyperref}
\usepackage{ifthen}

\makeatletter
\renewcommand{\@biblabel}[1]{}
\renewcommand{\@cite}[2]{%
{#1\ifthenelse{\boolean{@tempswa}}{,#2}{}}}
\makeatother

\hypersetup{breaklinks=true
,colorlinks=true,linkcolor=black,urlcolor=blue
,citecolor=black}

\ofoot{\thepage}
\ifoot{}

\setheadsepline{1pt}

\setkomafont{pagehead}{\normalfont\sffamily}
\setkomafont{pagenumber}{\normalfont\rmfamily}

\usepackage{booktabs}
\usepackage{amsmath}
\usepackage{amssymb}
\usepackage{multicol}
\usepackage{float}

\makeatletter
\newcommand{\listofcontributions}{\@starttoc{con}}

\newcommand{\l@contribution} {\@dottedtocline{1}{1.5em}{2.3em}}
\makeatother

\newenvironment{contribution}{
\setcounter{section}{0}
\setcounter{figure}{0}
\setcounter{table}{0}
\begin{flushleft}
{\em Clumping in Hot Star Winds \\
W.-R.\ Hamann, A.\ Feldmeier \& L.\ Oskinova, eds.\\
Potsdam: Univ.-Verl., 2007 \\
URN: http://nbn-resolving.de/urn:nbn:de:kobv:517-opus-13981
} 
\end{flushleft}
}{
\newpage
\lehead{}
\rohead{}
}

\begin{document}

\setlength{\baselineskip}{2.5ex}

\begin{contribution}

\newcommand{\teff}{{$T_{\rm eff}$}}
\newcommand{\logg}{{$\log g$}}
\newcommand{\iue}{{\it IUE}}
\newcommand{\fuse}{{\it FUSE}}
\newcommand{\ha}{H$_\alpha$}
\newcommand{\mdot}{$\dot{M}$}
\newcommand{\mdotq}{$\dot{M} q$}
\newcommand{\tradr}{$\tau_{rad}^R$}
\newcommand{\tradb}{$\tau_{rad}^B$}
\newcommand{\kc}{\kappa_C}
\newcommand{\kf}{\kappa_f}
\newcommand{\keff}{\kappa_{eff}}
\newcommand{\tc}{\tau_C}

\lehead{D.\ Massa, R. K. Prinja \& A. W. Fullerton}

\rohead{The effects of clumping on wind line variability}

\begin{center}
{\LARGE \bf The effects of clumping on wind line variability}\\
\medskip

{\it\bf D. Massa$^1$, R.K.\ Prinja$^2$ \& A.W.\ Fullerton$^3$}\\

{\it $^1$SGT, Inc., USA}\\
{\it $^2$University College London, England}\\
{\it $^3$Space Telescope SciInstitute, USA}

\begin{abstract}

We review the effects of clumping on the profiles of resonance doublets. By 
allowing the ratio of the doublet oscillator strenghts to be a free 
parameter, we demonstrate that doublet profiles contain more information 
than is normally utilized.  In clumped (or porous) winds, this ratio can 
lies between unity and the ratio of the $f$-values, and can change as a 
function of velocity and time, depending on the fraction of the stellar 
disk that is covered by material moving at a particular velocity at a given 
moment. Using these insights, we present the results of SEI modeling of a 
sample of B supergiants, $\zeta$~Pup and a time series for a star whose 
terminal velocity is low enough to make the components of its Si~{\sc 
vi}$\lambda\lambda1400$ independent.  These results are interpreted within 
the framework of the Oskinova et al.~(\cite{massa:Oskinova}) model, and 
demonstrate how the doublet profiles can be used to extract infromation 
about wind structure.  

\end{abstract}
\end{center}

\begin{multicols}{2}

\section{Introduction}

Clumping has been apparent from time series for many years (Kaper et al.\ 
\cite{massa:Kaper}, Prinja et al.\ \cite{massa:Prinja02}).  In this 
contribution we will first show how clumping affects doublet ratios and 
then use a time series to demonstrate the presence of clumping in winds.  

\section{Clumping}

It is well known in the AGN community that doublet ratios are sensitive to 
clumping, with the ratio related to the covering factor of the clumped 
medium (e.g., Ganguly et al.\ \cite{massa:Ganguly}).  For an extended 
source, the optical depth ratio determined from a doublet lies between the 
ratio of the blue and red $f$-values ($\alpha \equiv f_B/f_R$) and unity, 
depending on the covering factor of the source.  Further, it is possible to 
interpret the observed doublet ratio within the framework of the Oskinova 
et al.\ (\cite{massa:Oskinova}) model.  In their model, the doublet 
components have an {\em observed} optical depth ratio of 
\begin{equation}
\frac{\tau_{rad}^B}{\tau_{rad}^R} = \frac{\keff^B}{\keff^R} = 
  \frac{1 -e^{-\tc^B}}{1 -e^{-\tc^B/\alpha}} \; ,
\label{massa:ratio} 
\end{equation}
which varies between 1 and $\alpha$ (see, Massa et al.\ 2003, for a
definition of $\tau_{rad}$, and Oskinova et al. for definitions of the 
other symbols).  Consequently, if we allow the ratio of $f$-values to be 
a free parameter, 
\begin{equation}
\frac{\tau_{rad}^B}{\tau_{rad}^R} = \frac{f_B^\prime}{f_R^\prime} = 
\frac{\keff^B}{\keff^R} \; .
\label{massa:primes}
\end{equation}
Thus, $f_B^\prime/f_R^\prime$ (determined by the fit) gives 
$\keff^B/\keff^R$ and, hence $\tc^B$.  Oskinova et al.\ also relate the 
measured and smoothed opacities (or optical depths) and the clump optical 
depths as follows: $\keff/\kf = (1 -e^{-\tc})/\tc$.  Since \mdotq\/ 
(where $\dot{M}$ is the mass loss rate and $q$ is the ionization fraction, 
see, e.g., Massa et al.\ \cite{massa:Massa}) should be derived from $\kf$, 
the observed \mdotq\/ must be corrected as follows
\begin{equation}
\dot{M} q =  (\dot{M} q)_{obs} \left(\frac{\tc}{1 -e^{-\tc}}\right)
\label{uvmdot}
\end{equation}
Fig.~\ref{massa:functions} shows how the ``observed ratio'' defines a point 
on the y-axis which gives $\tc^B$.  $\tc^B$ then gives the ratio of the 
effective to actual opacity, $\keff^B/\kf^B$.  Note: {\em to obtain 
$\keff^B \ll \kf^B$ and \mdot's near expected values}, requires $\tc^B 
\gtrsim 5$, which implies $\keff^B/\keff^R \lesssim 1.1$.  

\begin{figure}[H]
\begin{center}
\includegraphics[width=0.8\columnwidth]{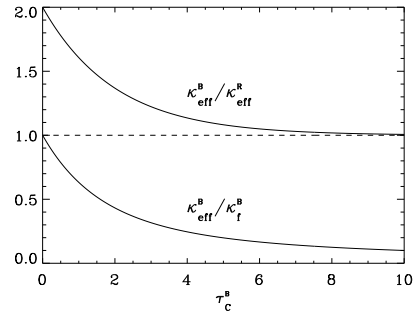}
\caption{Ratio of effective opacities, $\keff^B/\keff^R$, (top) and 
effective to smooth opacities, $\keff^B/\kf^B$, (bottom) versus 
clump optical depth, $\tau^B_C$, for a doublet with $\alpha = 2$.
\label{massa:functions}}
\end{center}
\end{figure}
\begin{figure}[H]
\begin{center}
\includegraphics[width=\columnwidth]{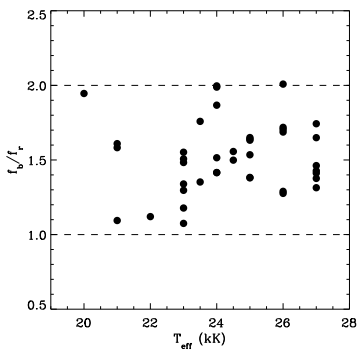}
\caption{Derived ratios of oscillator strengths versus $T_{\rm eff}$ for 
Si~{\sc iv} $\lambda\lambda1400$ in B supergiants with wind lines $0.3 
\leq \tau_{rad} \leq 5$ (weaker have inadequate optical depth information 
and stronger are too saturated).
\label{massa:bsupers}}
\end{center}
\end{figure}
\begin{figure}[H]
\begin{center}
\includegraphics[width=\columnwidth]{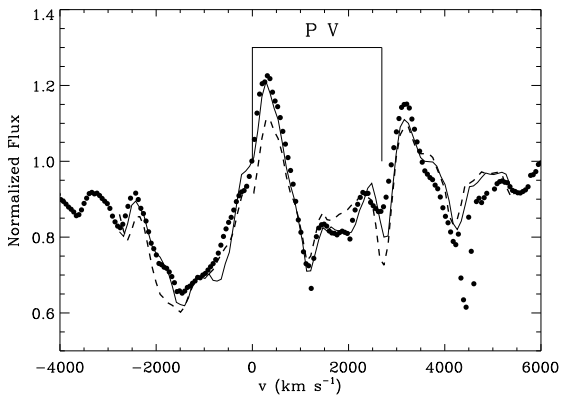}
\caption{P~{\sc v} in $\zeta$~Pup.  Points are data, dashed curve is SEI 
fit with fixed $f$-values, solid curve allowed the ratio of $f$-values to 
vary.\label{massa:zet_pup}}
\end{center}
\end{figure}

It must now be determined whether this additional information is present in 
the observed profiles.  As a first test, we fit the Si~{\sc iv}$\lambda
\lambda$1400 doublet in the sample of B supergiants given in Prinja et al.\ 
(\cite{massa:Prinja05}) using a variable $f$-value.  The results are shown 
in Figure~\ref{massa:bsupers}.  Notice that all of the $f$-value ratios lie 
between 1 and 2, as expected for a clumped wind.  

As a second test, we fit $Copernicus$\ data of the P~{\sc v}$\lambda\lambda 
1117, 1128$ doublet in $\zeta$~Pup.  Two least squares SEI (Lamers et al.\ 
\cite{massa:Lamers}) fits are shown in Fig.~\ref{massa:zet_pup}.  Both use 
a $T_{eff} = 40$kK, $\log g = 3.5$ TLUSTY model photosphere.  The dashed 
fit uses the actual ratio of $f$-values, 2.02.  The solid fit varied the 
ratio, giving a best fit value of 1.84.  This 10\% change clearly improves 
the fit, making the blue absorption weaker relative to the red.  The fit 
gives $\tau_C^B \simeq 0.5$ and $\keff^B/\kf^B \simeq 0.5$. Thus, the 
correction to \mdotq\ is only 1.3, increasing ratio of P~{\sc v} to radio 
mass loss rates given by Fullerton et al.\ (\cite{massa:Fullerton}) 
from 0.11 to 0.14 -- {\em still far smaller than expected}, even if 
$q($P~{\sc v}$) \sim 0.5$.  Thus the mass loss rate in this case is truly 
smaller than expected.  

\begin{figure}[H]
\includegraphics[width=0.4\textwidth, angle=90]{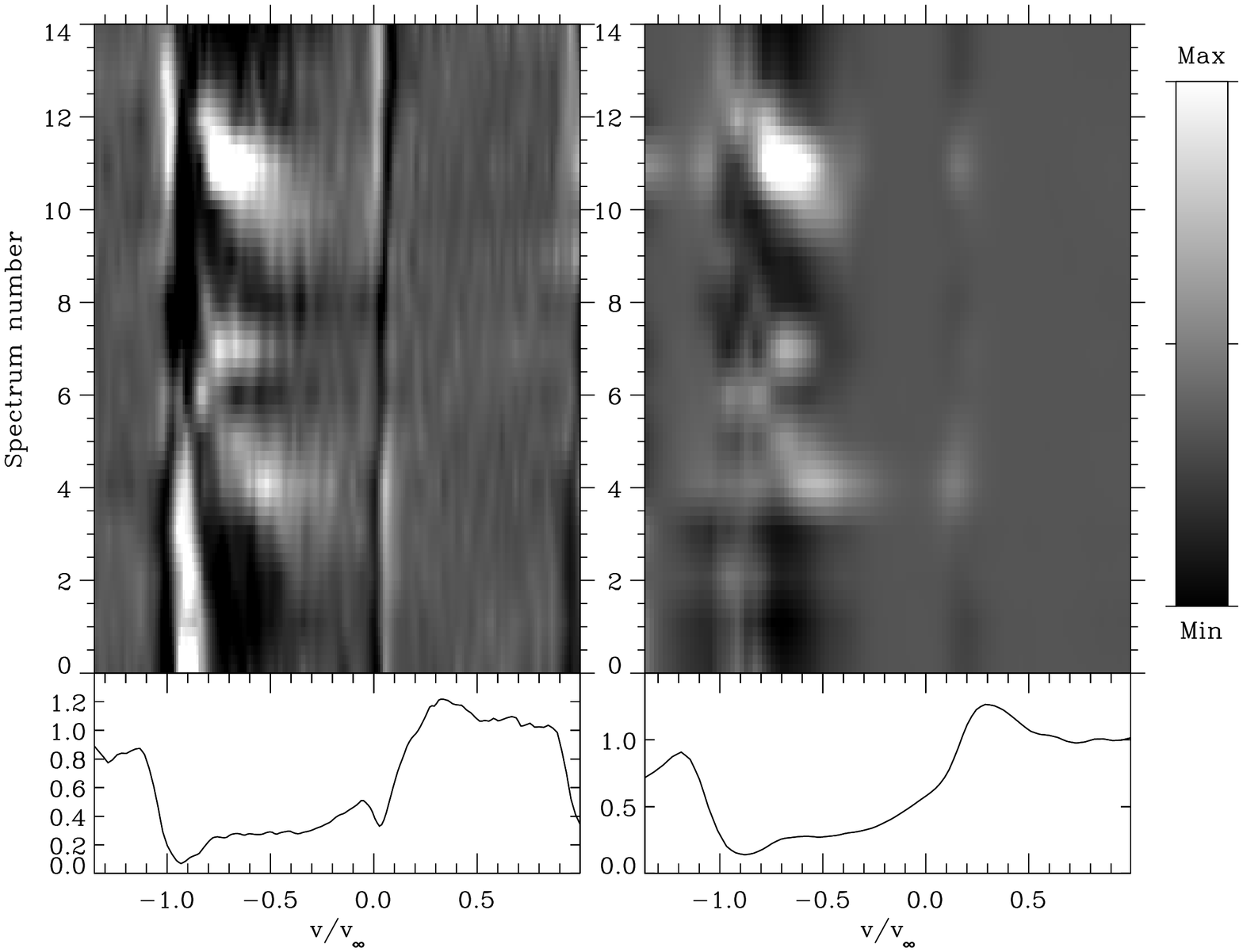}
\hfill\includegraphics[width=0.4\textwidth, angle=90]{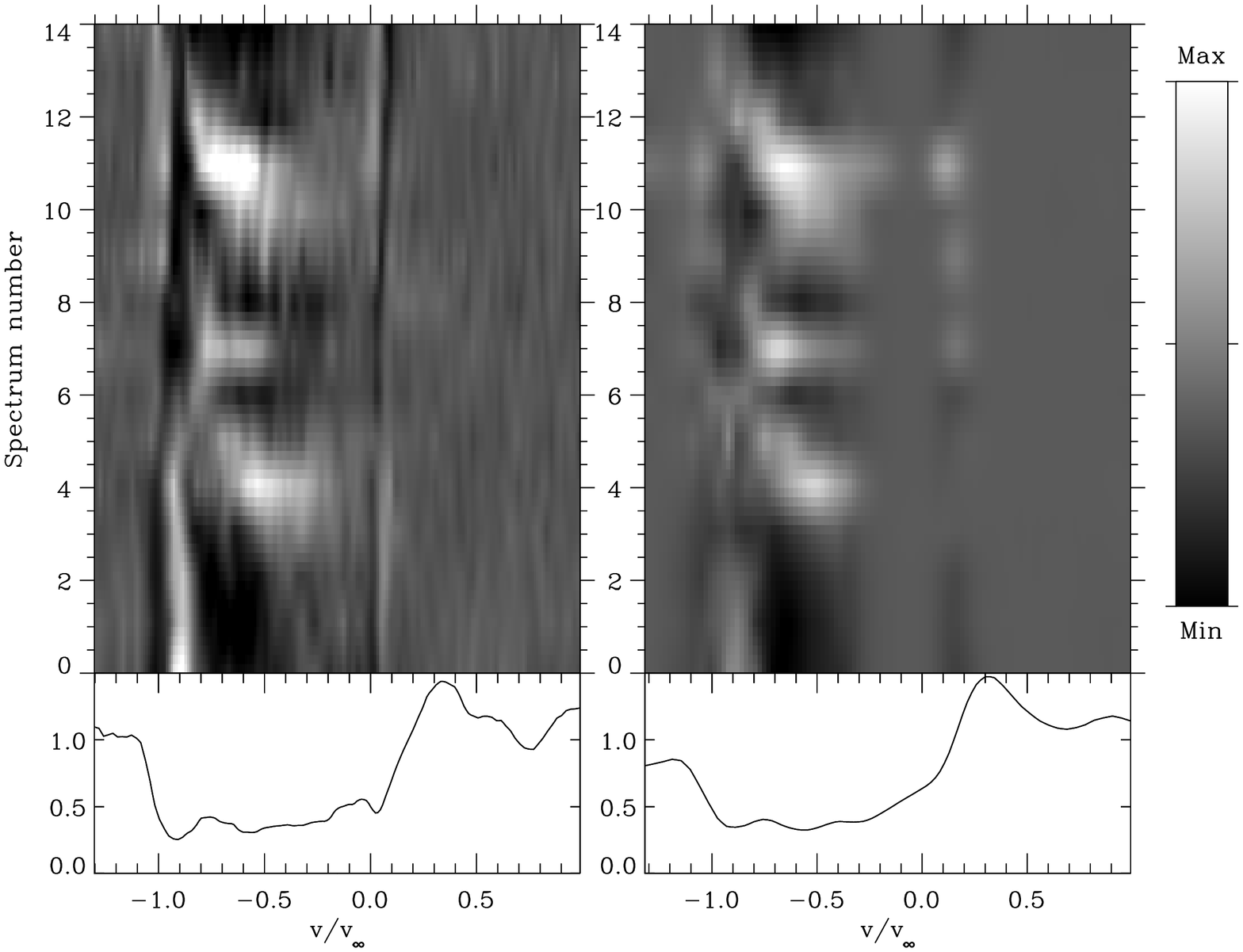}
\caption{Dynamic spectra of the observed (left) and modeled (right) spectra  
of the blue (top) and red (bottom) components of Si~{\sc iv}$\lambda
\lambda1400$ in HD 47240.  The spectra are normalized by their means.
\label{massa:dynamic}}
\end{figure}

\section{Clumping and time series}

\begin{figure*}[!t]
\begin{center}
\includegraphics
  [width=0.5\textwidth, angle=90]{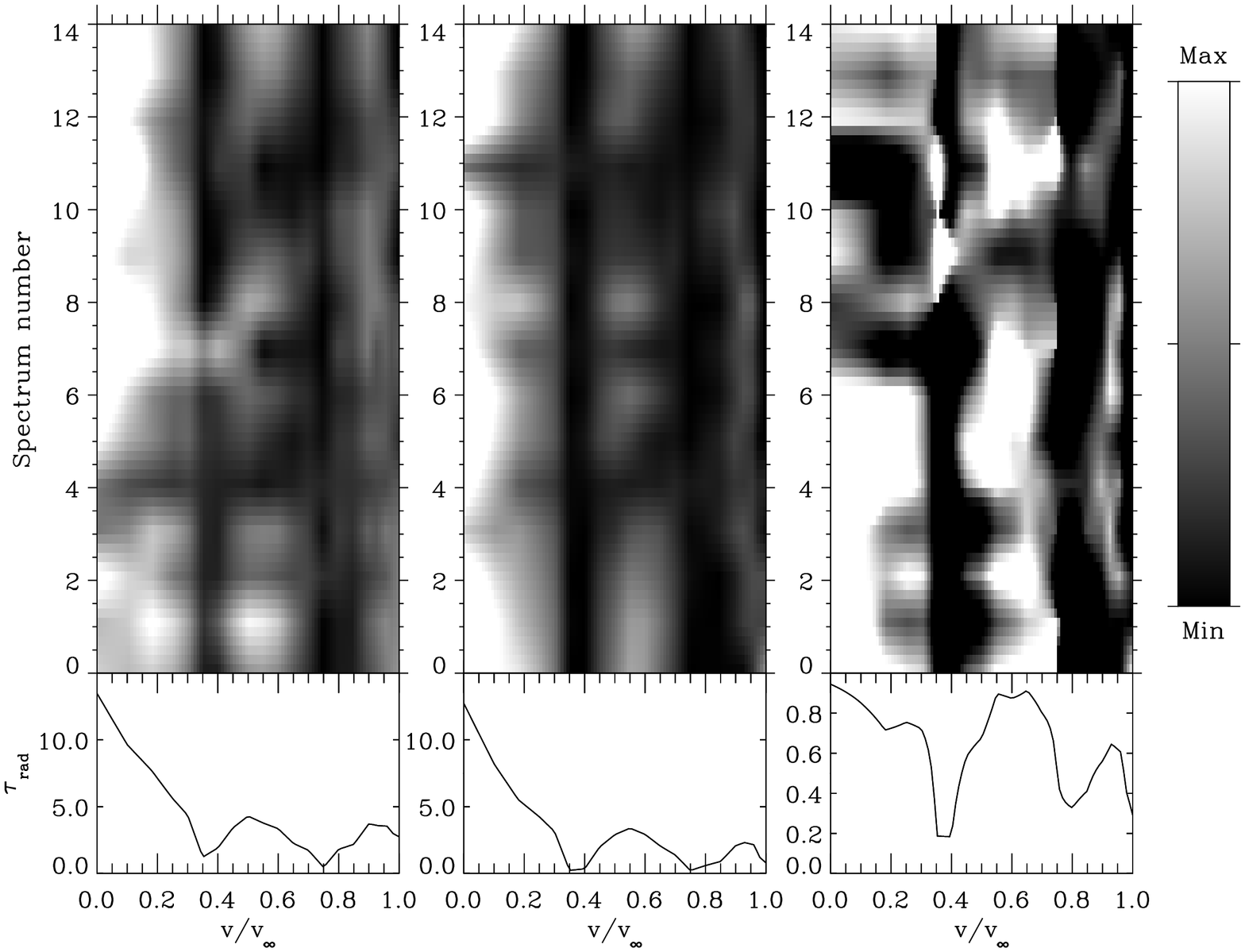}
\caption{Dynamic spectra for \tradb\ (left), \tradr\ (middle) and 
\tradr/\tradb\ (right) in HD~47240.  
\label{massa:taus}}
\end{center}
\end{figure*}

Time series for stars with $v_\infty \geq 2 c \Delta \lambda/\lambda$, 
where $\Delta \lambda$ is the doublet separation, present a valuable test 
for the effects of clumping on wind lines.  The B1 Ib, HD 47240, is such a 
star with $v_\infty = 980$ km s$^{-1}$.  Consequently, each of its Si~{iv} 
components are effectively independent.  Thus, they were fit independently.  
The result is equivalent to allowing $f_b/f_r$ to be a function of 
velocity.  The results are shown in Fig.~\ref{massa:dynamic}, which shows 
the observed and modeled normalized dynamic spectra versus spectrum number 
for each component, and Fig.~\ref{massa:taus}, which shows the un-normalized 
optical depths and their ratio.  Several points are noteworthy.  First, 
{\em both} components show similar, persistent velocity dependent 
structure.  The fits of many B supergiants given by Prinja et al.\ 
(\cite{massa:Prinja05}) had similar structure appearing in different ions 
in the same star.  Thus, this structure may be real.  Whether it results 
from density inhomogeneties or velocity plateaus, cannot be determined.  
Second, when $\tau_{rad}$\ is greater than a few tenths and well-defined, 
the ratio of optical depths varies between 1 (clumped) and 0.5 (unclumped).  
Third, it appears that the ratio (clumping) decreases at velocities where 
the density decreases.  Fourth, there seems to be a general decrease in the 
clumping at larger velocity, similar that seen in O stars by Puls et al.\ 
(\cite{massa:Puls}).

\section{Discussion}

It has been demonstrated that resonance doublets contain more information 
than is usually exploited, that this information is related to clumping, 
and that it can be interpreted by the Oskinova et al.\ (\cite{massa:Oskinova}) 
model.  Our ultimate goal is to apply these techniques to a large number of 
stars with a range of stellar and wind parameters.  This will allow us to 
examine how these relate to the empirically determined clumping parameters 
and provide clues to the physical agents responsible for wind clumping.

\end{multicols}

\end{contribution}


\end{document}